\documentclass[11pt]{article}

\usepackage{amsmath}    
\usepackage{graphics,graphicx}   
\usepackage{verbatim}   
\usepackage{color}      
\usepackage{subfigure}  
\usepackage{hyperref}   
\usepackage{lineno}
\usepackage{amssymb}

\newcommand*\farcm{\ensuremath{\overset{\prime}{.}}}
\newcommand*\farcs{\ensuremath{\overset{\prime\prime}{.}}}

\begin{document}

\begin{center}
{\large\bf 
The inside-out planetary nebula around a born-again star
} 
\end{center}

\noindent
{\bf 
Mart\'\i n A.\ Guerrero$^1$, 
Xuan Fang$^2$, 
Marcelo M.\ Miller Bertolami$^{3,4}$, 
Gerardo Ramos-Larios$^5$, 
Helge Todt$^6$, 
Alexandre Alarie$^7$, 
Laurence Sabin$^7$, 
Luis F.\ Miranda$^1$, 
Christophe Morisset$^7$,
Carolina Kehrig$^1$, and
Sa\'ul A.\ Zavala$^8$
}
\\
\\
$^1$ Instituto de Astrof\'\i sica de Andaluc\'\i a, IAA-CSIC, Granada, Spain \\
$^2$ Laboratory for Space Research \& Department of Physics, Faculty of 
Science, The University of Hong Kong, Hong Kong, People's Republic of China \\
$^3$ Instituto de Astrof\'\i sica de La Plata, UNLP-CONICET, La Plata, Argentina \\
$^4$ Facultad de Ciencias Astron\'omicas y Geof\'\i sicas, UNLP, La Plata, Argentina \\
$^5$ Instituto de Astronom\'\i a y Meteorolog\'\i a (IAM), Dept.\ de F\'\i sica, CUCEI, Universidad de Guadalajara, Guadalajara, Jalisco, Mexico \\
$^6$ Institute of Physics and Astronomy, University of Potsdam, Potsdam, Germany \\
$^7$ Instituto de Astronom\'\i a, Universidad Nacional Aut\'onoma de M\'exico, Ensenada, B.C., Mexico \\
$^8$ Tecnol\'ogico Nacional de M\'exico/I.T.\ Ensenada, Departamento
de Ciencias B\'asicas, Ensenada, B.C., Mexico.


\noindent
{\bf 
Planetary nebulae are ionized clouds of gas formed by the hydrogen-rich 
envelopes of low- and intermediate-mass stars ejected at late evolutionary 
stages. 
The strong UV flux from their central stars causes a highly stratified 
ionization structure, with species of higher ionization potential closer 
to the star.  
Here we report on the exceptional case of HuBi\,1, a double-shell planetary 
nebula whose inner shell presents emission from low-ionization species close 
to the star and emission from high-ionization species farther away. 
Spectral analysis demonstrates that the inner shell of
HuBi\,1 is excited by shocks, whereas its outer shell 
is recombining.  
The anomalous excitation of these shells can be traced to its 
low-temperature [WC10] central star whose optical brightness 
has declined continuously by 10 magnitudes in a period of 46 years.
Evolutionary models reveal that this star is the descendent of 
a low-mass star ($\simeq$~1.1 $M_\odot$) that has experienced 
a born-again event\cite{DLSetal2000} whose ejecta shock-excite 
the inner shell. 
HuBi\,1 represents the missing link in the formation of metal-rich 
central stars of planetary nebulae from low-mass progenitors, offering 
unique insight regarding the future evolution of the born-again Sakurai's 
object\cite{HZHetal2005}. 
Coming from a solar-mass progenitor, HuBi\,1 represents  
a potential end-state for our Sun.  
}

\noindent
Planetary nebulae (PNe) are a short-lived $\approx$20,000 yr period in the 
transition of low- and intermediate-mass stars ($M_{initial}=0.8-8.0 M_\odot$) 
from the Asymptotic Giant Branch (AGB) phase towards the white-dwarf (WD) 
phase.
The ionization structure of PNe, governed by the distance of the nebular 
material to the central star (CSPN), is well-known to display an “onion-like”
structure with higher ionization species such as He$^{++}$ and O$^{++}$ close
to the central star, lower ionization species such as N$^+$ and O$^+$ in the
outer region, and neutral and molecular species such as O$^0$ and H$_2$ in
the outermost photo-dissociation region.

\noindent
Whereas this is the general rule, we have discovered an exceptional 
case in HuBi\,1\cite{HB1990} (PN~G012.2+04.9, a.k.a.\ PM\,1-188).
Originally reported to have a faint bipolar outer shell and an unresolved 
bright inner shell\cite{PH1994}, our sub-arcsec resolution images (Figure~1a) 
and spatial profiles of selected emission lines (Figure~1b-d) extracted from 
sub-arcsec long-slit spectroscopic observations refine this description, while 
unveiling a puzzling fact: the outer shell of HuBi\,1, a barrel-like structure 
with faint polar protrusions slightly inclined to the line of sight (see 
Methods), surrounds a [N~{\sc ii}]-bright inner shell (Figure~1b).
The low-excitation inner shell of HuBi\,1 is unusual among PNe,
but its ionization structure is even more peculiar (Figure 1c,d). 
The spatial profile of the [O~{\sc ii}] $\lambda$3727 emission line
fills the inner shell cavity whose rim is defined by the [N~{\sc ii}]
$\lambda$6548 and [O~{\sc iii}] $\lambda$5007 emission lines. 
This is exactly opposite to typical photo-ionized nebulae, where 
the [N~{\sc ii}] and [O~{\sc ii}] spatial profiles are generally
coincident with each other and the [O~{\sc iii}] profile peaks inside, 
given the very similar ionization potentials required for O$^+$ and 
N$^+$ and the higher ionization potential for O$^{++}$.  
More surprisingly, perhaps, the spatial profile of the He~{\sc ii} 
$\lambda$4686 line peaks outside those of [N~{\sc ii}] and [O~{\sc iii}], 
although the ionization potential for He$^{++}$ is higher than for 
N$^+$ and O$^{++}$. 
The ionization structure of the inner shell of HuBi\,1 is 
inverted with respect to that of photo-ionized nebulae.
In this sense, HuBi\,1 is inside-out.

\noindent
The origin of the inverted ionization structure of the inner shell of 
HuBi\,1 must be at its CSPN, IRAS\,17514-1555 (hereafter IRAS\,17514).  
This star is of [WR]-type, i.e.\ its spectrum shows relatively 
broad emission lines similar to those of massive Wolf-Rayet (WR)
stars\cite{H1996} that indicate H-deficient strong stellar winds. 
The spectrum of IRAS\,17514, which reveals a wealth of C~{\sc ii} and
C~{\sc iii} emission lines\cite{PH1994} arising from a C-rich stellar 
wind, has been assigned a spectral type of [WC10]\cite{P2005}.
Our analysis of mid-1990s spectroscopic observations of IRAS\,17514 using 
state-of-the-art non-LTE models (see Methods) confirms its low terminal 
wind velocity $v_\infty\approx360$ km~s$^{-1}$, low surface temperature 
$T_{\star}\approx38,000$ K, and high C and O abundances\cite{LH1998}.  
The number of He$^+$ ionizing photons suggested by this non-LTE stellar model,
$\log Q$(He~{\sc ii})$<$34, is much lower than that derived for the nebula,
$\log Q$(He~{\sc ii})$>$43.8 (see Methods).  
Not only is the spatial distribution of the He~{\sc ii} emission in HuBi\,1 
unexpected for a photo-ionized nebula, but its detection is puzzling because 
the CSPN is not hot enough to photo-ionize He$^+$.

\noindent
The actual state of affairs is more complex yet.  
A revision of long-term optical photometry and spectroscopy of IRAS\,17514 
(see Methods) reveals it has faded continuously over time (Figure~2). 
Back to March 1989, the star had a $V$-band magnitude of 14.6, which 
faded to 19.8 in June 2014, and was above a detection limit $\geq$22.7 
in May 2017. 
Using archival USNO~B-1 data obtained on January 1971, we found 
a decline in the $B$ and $R$ bands by nearly 10 mags, implying a 
decrease in optical brightness $\cong$10,000 times in 46 years.  
The spectral changes in IRAS\,17514 during the period from 1996 to 2014 are 
mainly restricted to its stellar continuum, as the profiles and equivalent 
widths of the spectral lines remained unchanged.  
Since spectral lines are very sensitive to $T_{\star}$, a change of 
$T_{\star} > 2,000$ K can be discarded.  
Such small variation in $T_{\star}$ would imply a tiny change in the optical 
flux, which depends almost linearly on $T_{\star}$ given that the optical 
region of a CSPN spectrum can be approximated with the Rayleigh-Jeans Law.  
Rather, the decline in time of the optical flux of IRAS\,17514 is associated
with a correlated increase in obscuration.

\noindent
At the present time, IRAS\,17514 cannot even provide the ionizing flux
required to keep hydrogen ionized in HuBi\,1, which is estimated to be
$\log Q$(H~{\sc i})=45.7. 
With such a rapidly fading CSPN, the nebula is expected to 
exhibit unusual spectral properties.  
Indeed, the spectra of the inner and outer shells have their own 
share of peculiarities (see Methods). 
The outer shell is dominated by H~{\sc i} and He~{\sc i} recombination lines, 
with much fainter emission of forbidden lines, which is reminiscent of
recombining haloes in PNe\cite{CSSetal2003}.  
MAPPINGS photo-ionization models following the time-evolution of a low-density 
photo-ionized PN after the ionizing UV flux from its CSPN ceases show that its 
cooling time-scale is only a few decades, whereas its recombination time-scale 
is much longer, several thousand years (see Methods). 
As for the inner shell, the brightest emission lines in its spectrum are 
those of [N~{\sc ii}], with the notable detection of the He~{\sc ii} line 
at 4686 \AA. 
The inverted ionization structure and spectrum of this shell are 
typical of shocks propagating through an ionized medium\cite{DM1993}.  
Indeed, where the CLOUDY photo-ionization models using the best-fit 
non-LTE stellar atmosphere model of IRAS\,17514 fail to reproduce 
those, MAPPINGS models including shock excitation are successful for 
shock velocities $\gtrsim$70 km~s$^{-1}$ (see Methods). 
As the shock travels outwards through the ionized outer shell of
HuBi\,1, He~{\sc ii} emission arises at the location of the shock, 
while the [O~{\sc ii}], [O~{\sc iii}], and [N~{\sc ii}] emissions 
at the post-shock cooling region.

\noindent
Therefore, HuBi\,1 is a fossil nebula surrounding a shock-excited inner shell. 
Its CSPN, IRAS\,17514, has recently started ejecting large 
amounts of carbon-rich material at speeds faster than the 
nebular expansion.
The expansion of this ejecta shock-excites the inner shell and, as it 
streams away, it cools down to conditions optimal for condensation of 
dust grains\cite{PGGetal2009}. 
The increasing optical depth of the dusty circumstellar cocoon
shields the UV photons from the CSPN, so that the outer nebula 
cools down and has started recombining.

\noindent
Obviously, this is not the classical behavior for a CSPN.  
%
Dramatic drops in luminosity have been observed in slow 
novae\cite{NMAetal2012} and R CrB stars\cite{C1996}, 
but the spectral properties and variation timescale 
of these sources differ from those of IRAS\,17514. 
The C-rich dense stellar winds of (truly massive) WC stars provide
suitable sites for dust production\cite{W1995}, either in episodic
WC dust makers associated with eccentric binary systems with 
colliding winds (CWB) or persistent WC dust makers (WCd).
%
We note that the photometric behavior of WCd stars, with short 
(days to weeks) episodes of diminished brightness up to a few 
magnitudes\cite{W2014} does not fit that of HuBi\,1.  

\noindent
The key to understanding HuBi\,1 is the [WC] nature of its CSPN.
[WC]-type central stars account for a non-negligible fraction 
($\sim$15\%) of CSPNe\cite{AN2003}, but their origin is still 
debated.  
It has been proposed that they form through a thermal pulse, either at the 
end of the AGB (a final thermal pulse, FTP), during the post-AGB evolution 
of the CSPN when H-burning is still active (a late thermal pulse, 
LTP\cite{B2001}), or early in the cooling phase of the WD evolution (a very 
late thermal pulse, VLTP\cite{S1979}). 
These different channels 
result in different size and density of the PN, as well as distinct 
chemical composition of the CSPN.
An FTP will result in a dense and dusty nebula with bright IR 
emission around the [WC] star, which has been suggested for 
the IR-[WC] class\cite{Z2001}.  
The old kinematical age and low density of the outer shell of HuBi\,1 
($\approx$9,000 years and $\approx$200 cm$^{-3}$, respectively, see Methods) 
rather favor a VLTP. 
Born-again events such as V605\,Aql, V4334\,Sgr (a.k.a., the 
Sakurai's object), and FG\,Sge imply notable stellar temperature 
variations in short (a few years\cite{DLSetal2000,HZHetal2005})
timescales and even faster drops in luminosity, within days or
weeks; but these are not observed in IRAS\,17514. 
\emph{Ad-hoc} late stellar evolution models following VLTP events of low-mass
progenitors reveal loops in the HR diagram resulting in extended phases of
stalled $T_{eff}$ (see Methods).
The observed properties of HuBi\,1 (kinematical age) and its CSPN
(luminosity, surface temperature in the 1996-2014 period, and stellar 
wind abundances) can be reproduced with a VLTP in a 1.1 $M_\odot$ 
progenitor (Figure 3).

\noindent
IRAS\,17514 is caught in a brief but key episode in the evolution of a [WC] 
star and the nebula around it, which can be described by a low-mass born-again 
event.
The ejecta associated with this VLTP is propagating a shock through 
the surrounding outer shell of HuBi\,1 resulting in an inner shell 
with an inverted ionization structure unequaled among PNe, while its 
kinetic energy and significant enrichment in metals provide a direct 
clue to understanding the high turbulence of PNe around [WC]-type 
nuclei\cite{AGGetal2002} 
and their higher C and N abundances\cite{GPMetal2013}.
IRAS\,17514 provides the missing link between born-again events in
low-mass stars like V4334\,Sgr\cite{HZHetal2005} and fully developed
[WC]-nuclei, leaving open the possibility for our Sun itself, once it 
has become a PN\cite{GZM2018}, to experience a born-again event and 
conclude its life as a H-poor CSPN.   
The rapid evolution of HuBi\,1 and its CSPN deserves close monitoring. 
The possible expansion of its inner shell, continuous dust production around
the CSPN or its surface temperature increase can provide crucial insights to
the formation of [WC] stars and deposition of energy and enriched material in
the surrounding PNe.

\begin{figure}[t!]
\begin{center}
\vspace*{-1.0cm}
\includegraphics[height=16cm]{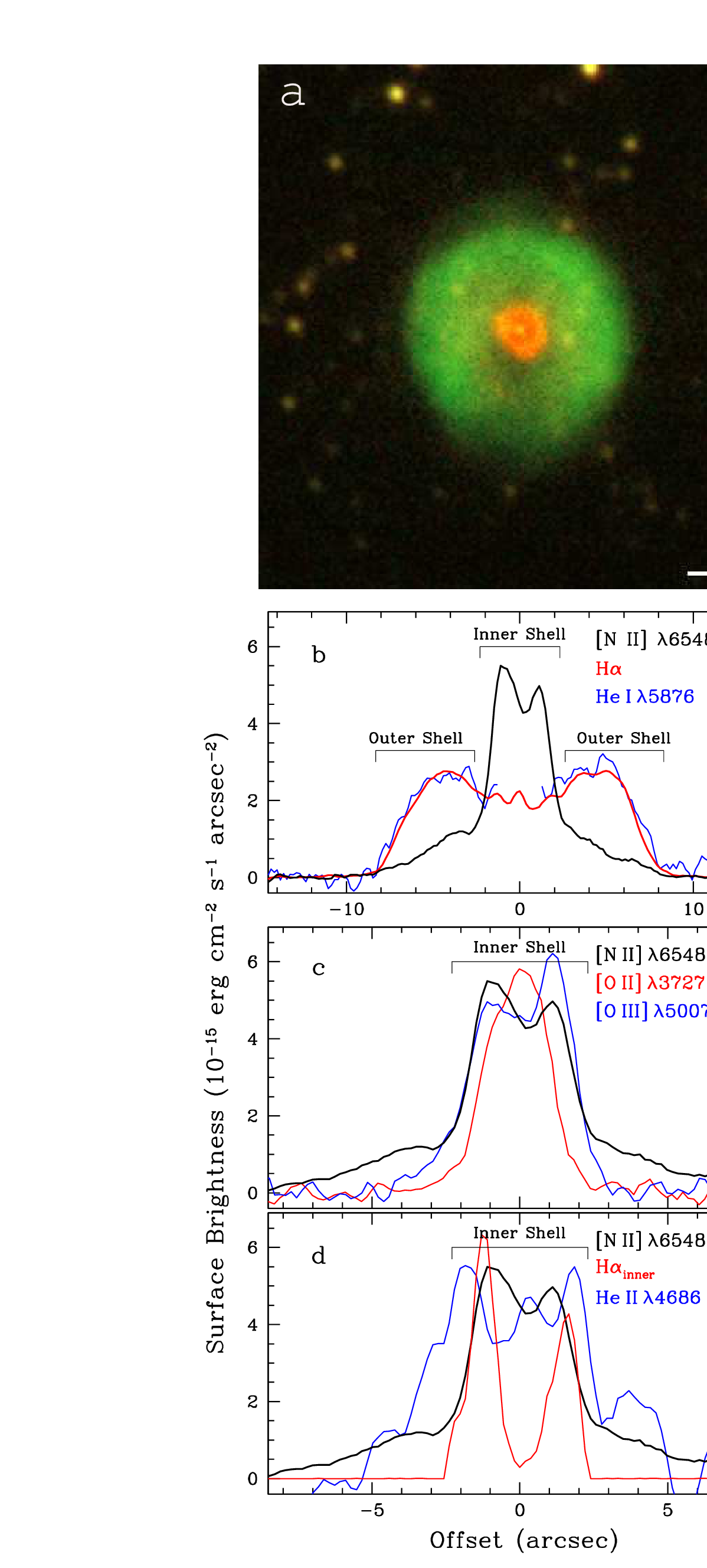}
\end{center}
\vspace*{-0.52cm}
{\small
{\bf
Figure 1.--
Color composite picture and spatial profiles of selected lines of HuBi\,1. }
(a) NOT ALFOSC [N~{\sc ii}] $\lambda$6584 (red) and H$\alpha$ $\lambda$6563 
(green) color composite picture of HuBi\,1.  
H$\alpha$ emission dominates the 18\farcs4$\times$19\farcs2 
outer shell, whereas the $\simeq$4$^{\prime\prime}$ inner shell is brighter 
in [N~{\sc ii}].  
(b-d) Continuum-subtracted spatial profiles of nebular emission lines
(see Methods) along the East-West direction of HuBi\,1 across its
central star for the outer (b) and inner shells (c-d).  
The H$\alpha$ spatial profile of the inner shell in panel \emph{d} is 
computed by subtracting to the total H$\alpha$ emission profile in 
panel \emph{b} a model for the emission from the outer shell (see Methods).  
The extent of the inner and outer shells, as used to extract
one-dimensional spectra, is marked.   
}
\vspace*{-0.55cm}
\end{figure}

\begin{figure}[t!]
\begin{center}
\includegraphics[height=12cm]{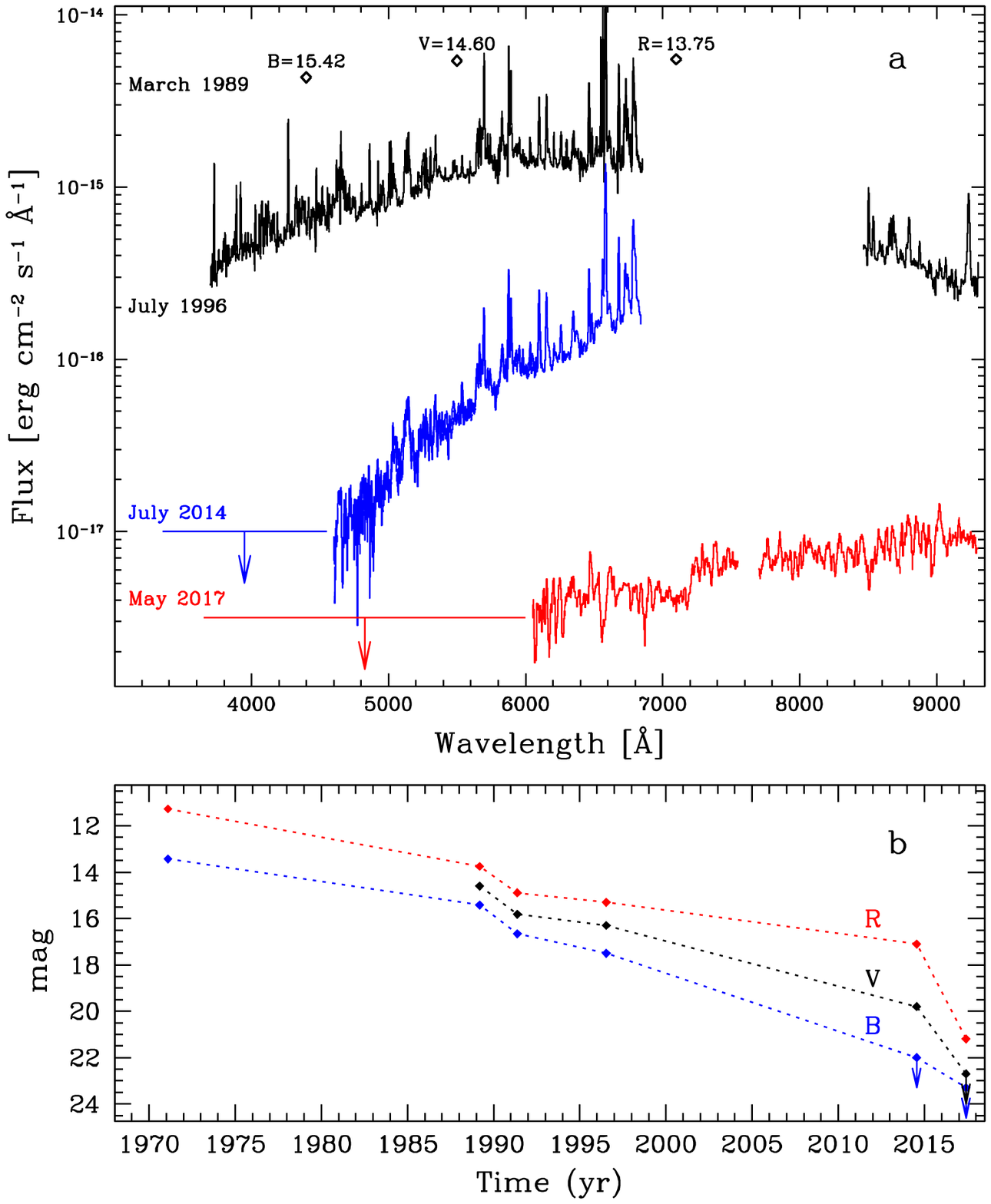}
\end{center}
\vspace*{-0.5cm}
{\small
{\bf
Figure 2.--
Long-term spectro-photometric evolution of IRAS\,17514-1555, the central
star of HuBi\,1.
}
(a) Optical spectra obtained at different epochs between July
1996 and May 2017.
The March 1989 $BVR$ magnitudes of the discovery paper of HuBi\,1\cite{HB1990}
are included for reference.  
The horizontal lines and down-pointing arrows indicate detection upper
limits.
(b) Time evolution of the $BVR$ magnitudes of IRAS\,17514
as derived from photometric or spectroscopic measurements.
}
\vspace*{-0.55cm}
\end{figure}

\begin{figure}[t!]
\begin{center}
\includegraphics[height=12cm]{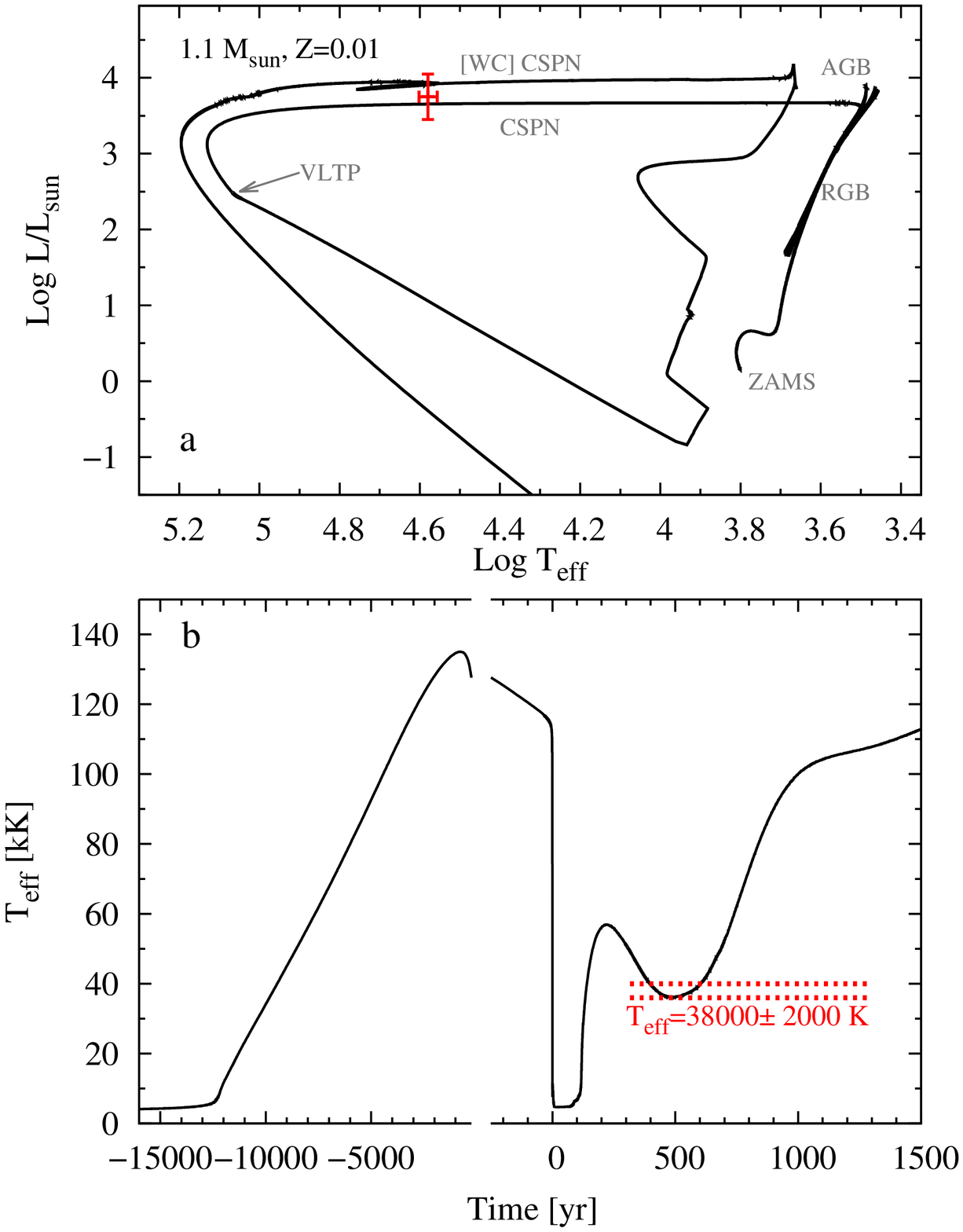}
\end{center}
\vspace*{-0.5cm}
{\small
{\bf
Figure 3.--
Evolutionary sequence of a PN progenitor with initial mass 1.1 $M_\odot$ that
experiences a VLTP }
(see Methods for additional details).  
(a) Evolutionary track in the HR diagram.  
The post-AGB star has a mass of 0.551 $M_\odot$. 
The red cross indicates the current location (with uncertainties) of HuBi\,1 
CSPN. 
(b) Post-AGB time evolution of $T_{eff}$.  
The time origin in this plot is set at the moment of the VLTP 
event, with different time scales being used before and after 
it.  
Note that the star departed from the AGB phase $\approx$12,000 yr before 
the occurrence of the VLTP, as marked by the initial increase of $T_{eff}$ 
in this plot.  
The red-dotted lines mark the uncertainty in the temperature 
of the central star of HuBi\,1, which is constrained by the 
ionization balance of the carbon and silicon ions as derived 
from the observed relative line strengths of C~{\sc ii}, 
C~{\sc iii}, and C~{\sc iv} lines, and Si~{\sc iii} and 
Si~{\sc iv} lines (see Methods for further details).  
}
\vspace*{-0.55cm}
\end{figure}

\clearpage

\noindent
{\LARGE\bf Methods}

\noindent
{\bf Optical image of HuBi\,1. }
Optical images of HuBi\,1 were obtained on September 2, 2008, using the
ALhambra Faint Object Spectrograph and Camera (ALFOSC) at the 2.5m Nordic
Optical Telescope (NOT) of the Observatorio de El Roque de los Muchachos
(ORM, La Palma, Spain).
The EEV 2K$\times$2K CCD camera was used, providing a pixel scale of
0.184$^{\prime\prime}$ pix$^{-1}$ and a field-of-view (FoV) of 6.3$^\prime$.  
Two 600s exposures were obtained through narrow-band filters that 
isolate the [N~{\sc ii}] $\lambda$6584 and H$\alpha$ $\lambda$6563 
emission lines.
These individual exposures were bias subtracted and flat-fielded using 
appropriate twilight sky frames, and then aligned and combined using 
standard {\sc iraf} routines.
The spatial resolution of the images, as derived from stars in the FoV,
is 0.65$^{\prime\prime}$. 
The images were combined into the color picture shown in Figure 1a to
highlight the different spatial locations of the H$\alpha$ recombination
and low-ionization [N~{\sc ii}] lines.

\noindent
{\bf Spatial profiles of selected emission lines of HuBi\,1. }
Long-slit optical spectra of HuBi\,1 were obtained on July 20, 2014 
using also ALFOSC at the 2.5m NOT telescope.
The 600 lines mm$^{-1}$ \#7 and \#14 grisms were used to acquire
two 1,200s exposures in the red and blue regions of the optical 
spectrum, respectively, covering the spectral range from 3250 to 
6840 \AA\ at a spectral resolution of 5.8 \AA.
The slit was placed at the central star along the East-West direction,
i.e. at a position angle (PA) of 90$^\circ$.
The spectro-photometric standard stars BD+33$^\circ$2642 and
BD+28$^\circ$4211 were used for flux calibration.
The seeing during the observations was 0.85$^{\prime\prime}$,
as determined from the FWHM of the continuum of field stars
covered by the slit.

\noindent
The two-dimensional spectra were used to extract spatial profiles of
emission lines of interest along the long-slit, i.e., along the East-West
direction across the central star.  
These spatial profiles were continuum subtracted using contiguous
spectral regions free from emission lines.
The emission line spatial profiles are 6 to 8 \AA\ wide,
whereas the redwards and bluewards contiguous background
spatial profiles add together a width of 48 \AA.  
Since the flux from the emission line is integrated and divided by the
long-slit width and pixel scale, the resulting spatial profile represents
the surface brightness of the nebula along this direction.
Note that the [N~{\sc ii}] $\lambda$6548 line is used for the
[N~{\sc ii}] spatial profile instead of the three times brighter
[N~{\sc ii}] $\lambda$6584 line because the latter is contaminated 
by bright stellar C~{\sc ii} $\lambda\lambda$6578,6582 emission 
lines.

\noindent
These profiles are shown in Figure 1b-d.
Panel \emph{b} shows the [N~{\sc ii}] $\lambda$6548, H$\alpha$
$\lambda$6563, and He~{\sc i} $\lambda$5876 emission lines, the 
latter multiplied by 20.
The spatial profiles of the H$\alpha$ and He~{\sc i} lines are similar
and reveal the location of the outer shell.
The spatial profile of the [N~{\sc ii}] line peaks inside and reveals
the location of the inner shell.
The inner shell is shown into more detail in panels
\emph{c} and \emph{d} of Figure 1.  
Panel \emph{d} reveals that the spatial profile of the He~{\sc ii}
$\lambda$4686 line extends outside that of the [N~{\sc ii}] $\lambda$6548 
line, whereas the H$\alpha$ inner shell synthetic profile peaks between 
those of He~{\sc ii} and [N~{\sc ii}]. 
This synthetic inner shell H$\alpha$ spatial profile was obtained after 
subtracting to the observed H$\alpha$ spatial profile in panel \emph{b} 
a two-Gaussian fit that represents the emission from the outer shell.

\noindent
    {\bf Age and morpho-kinematic modeling of the outer shell of
      HuBi\,1.}  
Long-slit, high-resolution spectra of HuBi\,1 were obtained with the 
Manchester Echelle Spectrograph (MES) attached to the 2.1m telescope 
of the OAN-SPM Observatory.
Four 1800s spectra were obtained on 2015 August 14 and 15 with the
slit oriented East-West (PA=90$^\circ$), and three 1800s spectra on
2018 May 5 with the slit oriented North-South (PA=0$^\circ$).  
In both cases, the slit was placed across the CSPN and the
detector was a 2K$\times$2K E2V CCD in a 2$\times$2 on-chip
binning, leading to a dispersion of 0.057 \AA~pix$^{-1}$ and 
a spatial scale of 0.35$^{\prime\prime}$~pix$^{-1}$.  
The slit length is 6\farcm5 and its width was set to 1$^{\prime\prime}$.  
A $\Delta\lambda=90$ \AA\ bandwidth filter was used to isolate the 87th 
order covering the H$\alpha$ and [N~{\sc ii}] $\lambda\lambda$6548,6584 
emission lines.  
A ThAr arc lamp was used for wavelength calibration to an accuracy of 1 
km~s$^{-1}$.  
The spectral resolution is 12 km~s$^{-1}$, as indicated by the $FWHM$ of 
the ThAr arc lines. 
The seeing was 1.2$^{\prime\prime}$--$1.5^{\prime\prime}$ during the observations.

\begin{figure}[h!]
\begin{center}
\includegraphics[height=7cm]{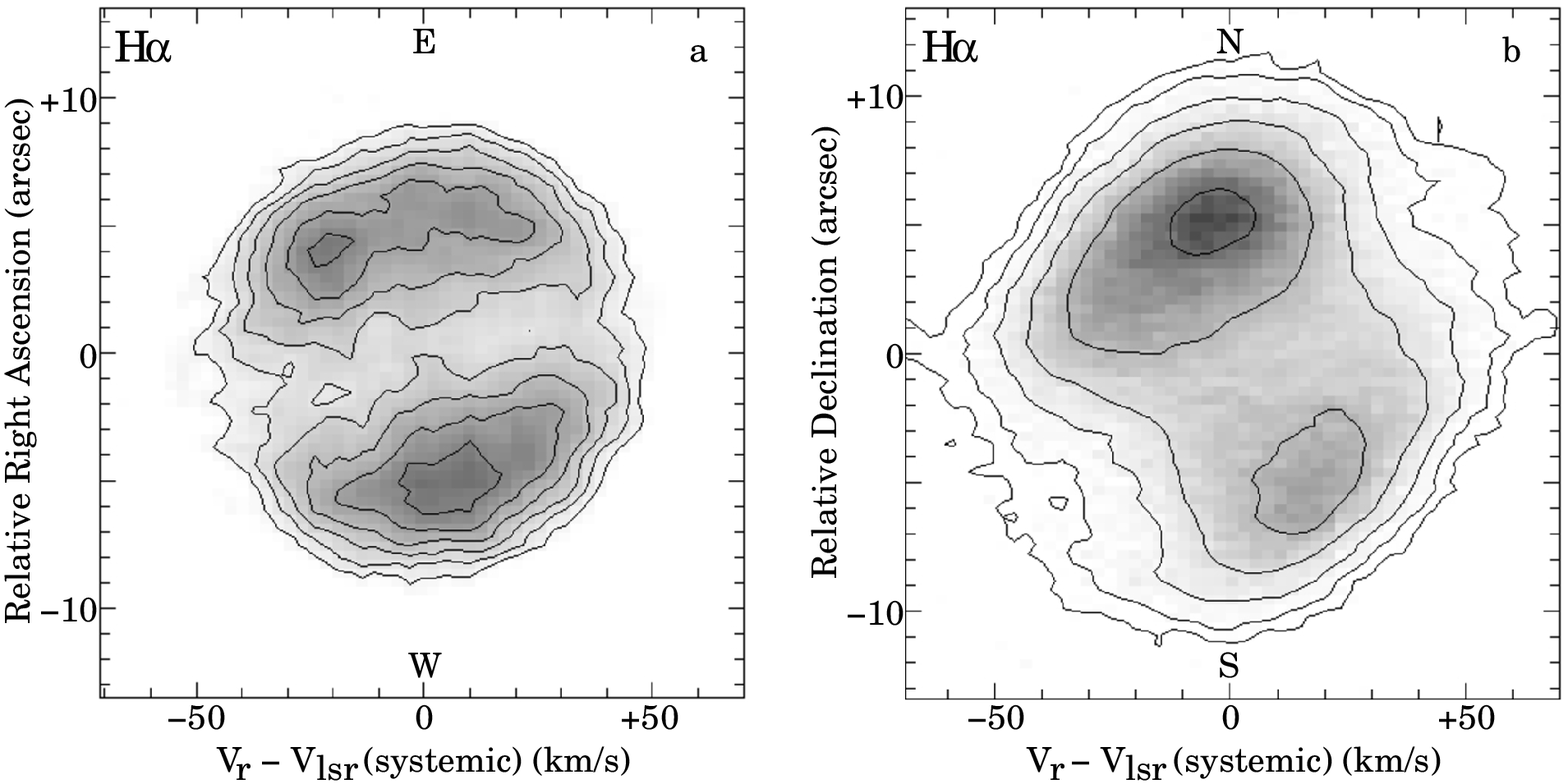}
\end{center}
\vspace*{-0.5cm}
{\small
{\bf
Supplementary Figure 1.--}
Position-velocity (PV) maps of the H$\alpha$ emission line from the outer
shell of HuBi\,1 as observed through a long-slit placed on the CSPN at PA 
90$^\circ$ (a) and PA 0$^\circ$ (b).
The line intensity is shown in grey-scale and contours.
The spatial origin corresponds to the position of the CSPN, whereas
the radial velocities are relative to the systemic velocity of the
nebula of +64.8 km~s$^{-1}$ as measured in the local standard of rest
(LSR).
}
\vspace*{-0.25cm}
\end{figure}

\noindent
The spectra at each PA were combined into a single long-slit spectrum. 
The position-velocity (PV) maps of the H$\alpha$ emission line
from the outer shell are shown in the Supplementary Figure~1.
The H$\alpha$ emission line in the PV map at PA 90$^\circ$ appears as a 
velocity ellipse, although the detailed distribution of the emission in 
this map, with faint emission at high velocity at the nebular center, is 
not completely consistent with a spherical shell.
Indeed, the open tilted line emission in the PV map at PA
0$^\circ$ is suggestive of an elongated open-ended structure
slightly tilted with respect to the line of sight.

\noindent
To reproduce the observed H$\alpha$ PV maps and image of HuBi\,1, 
we have used the three-dimensional morpho-kinematic code 
SHAPE\cite{Steffen_etal2011}. 
A barrel-like structure with fainter polar protrusions tilted with respect 
to the line of sight has been adopted to reproduce the general morphologies 
of the H$\alpha$ PV maps and image.  
A satisfactory fit is achieved with a barrel-like structure 
with aspect ratio $\gtrsim1.0$ whose symmetry axis is tilted 
to the line of sight by $\approx$25$^\circ$.  
Adopting a distance of 5.3 kpc\cite{FPB2016}, the kinematical 
age of the outer shell of HuBi\,1 is found to be $\approx$9,000 
yr.

\noindent
{\bf Non-LTE model of IRAS\,17514. }
Archival Isaac Newton Telescope (INT) Intermediate-Dispersion Spectrograph
(IDS) optical spectra of IRAS\,17514 (July 1996) were analyzed using the
Potsdam Wolf-Rayet (PoWR) model atmosphere code.
PoWR is a state-of-the-art non-LTE radiative transfer code that accounts
for mass-loss, line-blanketing, and wind clumping\cite{HG2004}.
It can be applied to a wide range of hot stars of arbitrary 
metallicities\cite{HPTetal2015,RRPetal2014} with different 
luminosity $L$, stellar temperature $T_\star$, surface gravity
$g_\star$, and mass-loss rate $\dot{M}$.

\begin{figure}[h!]
\begin{center}
\includegraphics[height=5cm]{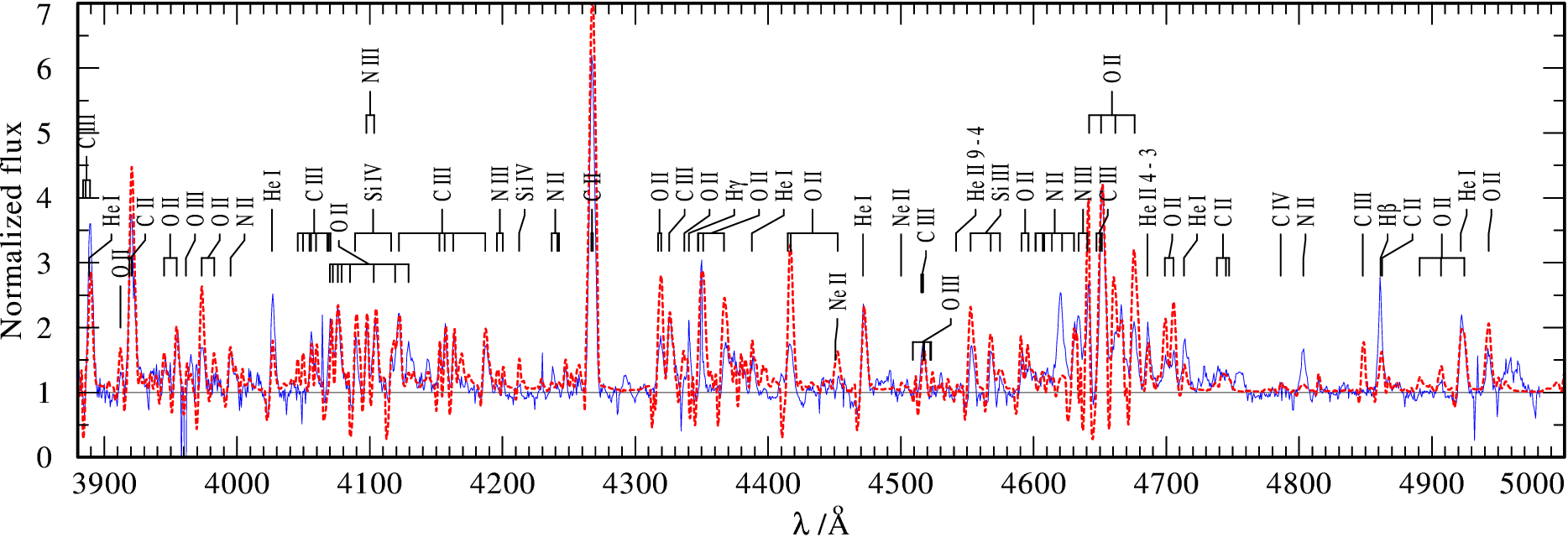}
\end{center}
\vspace*{-0.5cm}
{\small
{\bf
Supplementary Figure 2.-- }
Optical spectrum of IRAS\,17514 (blue) superimposed by the best-fit 
PoWR model (red). 
The identifications of key lines are labeled.
}
\vspace*{-0.25cm}
\end{figure}

\noindent
Initial stellar parameters\cite{LH1998} were refined on the basis of improved 
atomic data and line-blanket models including complex model atoms for H, He, 
C, N, O, Ne, Si and the iron group elements Sc, Ti, V, Cr, Mn, Fe, Co, and Ni. 
A temperature of 38,000$\pm$2,000 K is determined from 
the observed relative line strengths of  
C~{\sc ii} $\lambda\lambda$4266.9,4267.9, 
C~{\sc iii} $\lambda$5695.9, and 
C~{\sc iv} $\lambda\lambda$5801.3,5812.0, 
and those of 
Si~{\sc iii} $\lambda\lambda$4552.6,4567.8,4574.8, and 
Si~{\sc iv} $\lambda\lambda$4088.8 4116.1.  
For a temperature of 36,000~K, the C~{\sc iii} and C~{\sc iv} lines 
become much weaker in the model than in the observation, and the 
predicted Si~{\sc iii} to Si~{\sc iv} line ratio becomes higher than 
observed.  
On the other hand, for a temperature of 40,000~K, the C~{\sc iii}, 
C~{\sc iv}, and Si~{\sc iv} lines in the model become too strong, 
whereas the C~{\sc ii} and Si~{\sc iii} lines appear too weak.  
In this best-fit model, shown in the Supplementary Figure~2, 
the luminosity $\log$($L/L_\odot$) is 3.8$\pm$0.3, 
the extinction $E_{B-V}$ 1.50$\pm$0.05 mag, 
the stellar radius 1.8 $R_\odot$, 
the terminal wind velocity $v_\infty$ 360$\pm$30 km~s$^{-1}$, and 
the mass-loss rate $\approx$8$\times$10$^{-7}$ $M_\odot$~yr$^{-1}$.  
The chemical abundances by mass fraction are 0.01--0.05
for hydrogen, 0.33$\pm$0.10 for helium, and 0.5$\pm$0.1
for carbon, whereas rough estimates of 0.01 for nitrogen,
0.1 for oxygen, 0.04 for neon, and 0.01 for silicon are 
obtained.
Solar abundances of 1.4$\times10^{-3}$ were adopted for iron.

\noindent
{\bf H~{\sc i} and He~{\sc ii} ionizing flux. }
The H~{\sc i} and He~{\sc ii} ionizing flux can be derived from the 
intrinsic luminosity of the H$\alpha$ $\lambda$6563 and He~{\sc ii} 
$\lambda$4686 emission lines using the relations: \\
$ Q(\mathrm{H~I}) = L_{\mathrm{H}\alpha}/[j(\mathrm{H}\alpha)/\alpha_\mathrm{B}(\mathrm{H~I})] $, \\
$ Q(\mathrm{He~II}) = L_{\mathrm{He~II}~\lambda4686}/[j(\mathrm{He~II}~\lambda4686)/\alpha_\mathrm{B}(\mathrm{He~II})] $, \\
assuming case B recombination and $T_e = 10,000$ K in the low
density case for the emission $j$ and recombination $\alpha$ 
coefficients.  
Assuming that the inner and outer shells are spherical and that the surface 
brightness profiles of the H$\alpha$ and He~{\sc ii} $\lambda$4686 
emission lines can be represented by those extracted along the East-West 
direction presented in Figure~1b and Figure~1d, the total flux corrected 
for reddening of these lines are 2.0$\times$10$^{-12}$ and 
$>$1.8$\times$10$^{-14}$ erg~cm$^{-2}$~s$^{-1}$, respectively.
For the reddening correction, the extinction correction derived for the 
outer shell from the H$\alpha$ to H$\beta$ line ratio ($c$(H$\beta$)=1.1)
was applied.  
Since the extinction towards the inner shell might be larger, as 
suggested by an H$\alpha$/H$\beta$ line ratio larger than that of 
the outer shell, although based on an uncertain estimate of the 
H$\beta$ and H$\alpha$ fluxes of the inner shell, the He~{\sc ii} 
flux of this shell should be regarded as a lower limit.

By adopting a distance to HuBi 1 of 5.3 Kpc\cite{FPB2016}, the H$\alpha$ 
and He~{\sc ii} $\lambda$4686 \AA\ fluxes imply luminosities of 
$L_{\mathrm{H}\alpha}=7.0\times×10^{33}$ erg~s$^{-1}$ and 
$L_{\mathrm{He~II}~\lambda4686}\ge6.2\times10^{31}$ erg~s$^{-1}$. 
The corresponding H~{\sc i} and He~{\sc ii} ionizing photon fluxes are 
$\log Q(\mathrm{H~I})=45.7$ and $\log Q(\mathrm{He~II})\ge43.8$, respectively.  
These ionizing fluxes can be compared to the output predicted by our 
non-LTE best-fit model to the spectrum of the CSPN. 
For the 1996 spectrum of the CSPN, this model predicts 
$\log Q(\mathrm{He~II})=34.3$, much smaller than the 
value derived from the observed He~{\sc ii} lines. 
This same model predicts $\log Q(\mathrm{H~I})=48$, but the 2017 spectrum 
suggests that the H~{\sc i} ionizing flux has declined at least 4 orders 
of magnitude to $\log Q(\mathrm{H~I})\approx44$, also below the observed 
H~{\sc i} ionizing flux.

\begin{figure}[t!]
\begin{center}
\includegraphics[bb=18 160 592 718,height=9cm]{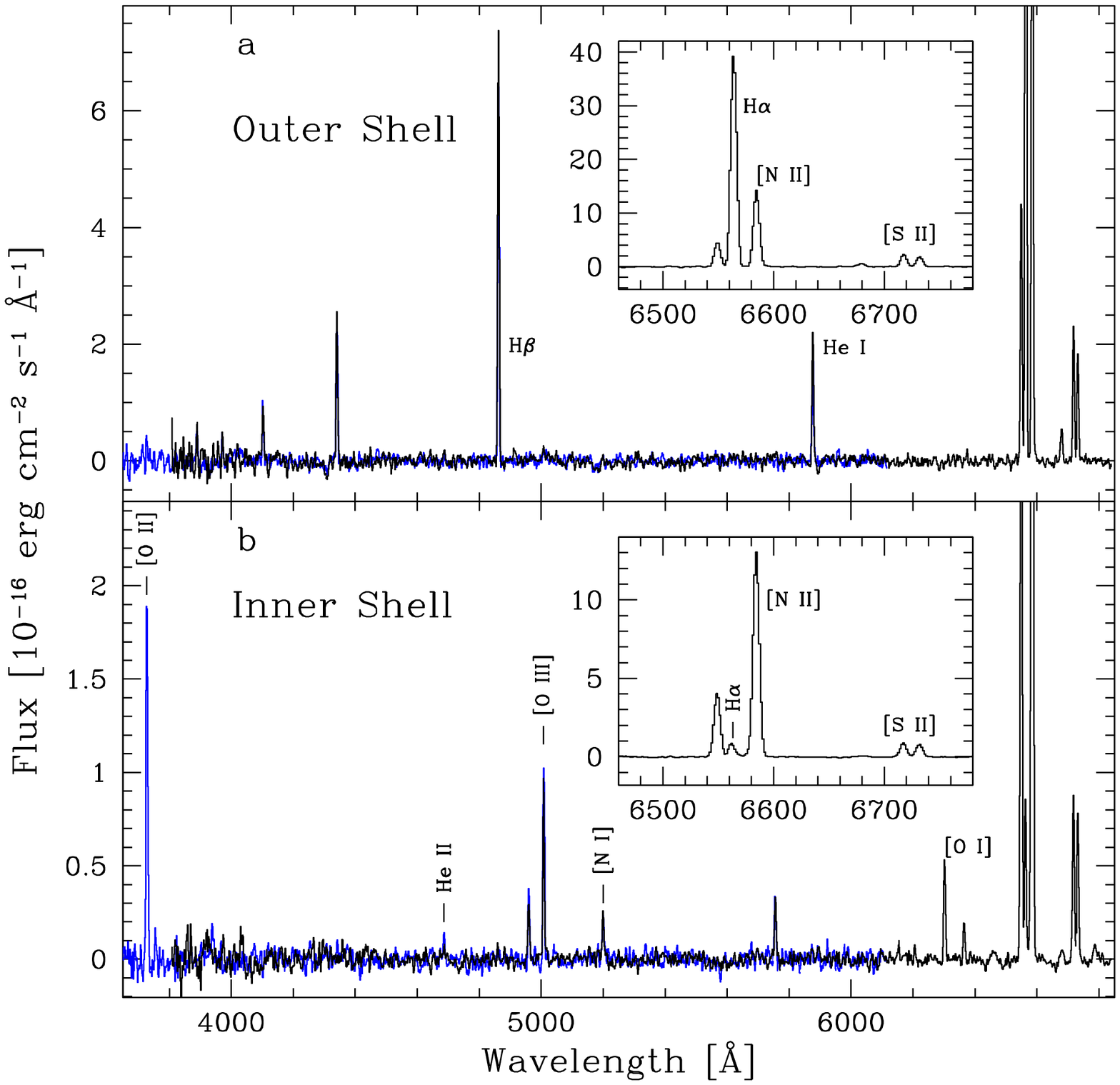}
\end{center}
\vspace*{-0.3cm}
{\small
{\bf
Supplementary Figure 3.-- }
NOT ALFOSC grism \#7 (black) and \#14 (blue) spectra of the outer (a) 
and inner (b) shells of HuBi\,1. 
The insets show the spectral range including the H$\alpha$, [N~{\sc ii}] 
$\lambda\lambda$6548,6584, and [S~{\sc ii}] $\lambda\lambda$6717,6731 
emission lines. 
Note the singular detection of the He~{\sc ii} $\lambda$4686 
emission line in the inner shell. 
}
\vspace*{-0.55cm}
\end{figure}

\noindent
{\bf Spectro-photometric evolution of IRAS\,17514.} 
May 1991\cite{PH1994}, July 1996 INT IDS, and July 2014 and May 2017 NOT
ALFOSC optical spectra have been used, in conjunction with the original
March 1989 $BVR$ photometric data\cite{HB1990} and January 1971 USNO B-1
Catalog $BR$ magnitudes\cite{MLCetal2003}, to investigate the
spectro-photometric time evolution of IRAS\,17514. 
The different spectra and photometric magnitudes are plotted in Figure~2b. 
The observed flux in each spectra has been corrected from the seeing 
and slit-width, although this correction is small, of a few percent. 
These spectra have also been used to fit a stellar continuum and 
determine the $BVR$ magnitudes of the central star at each epoch. 
The time evolution of the flux of the star at different photometric 
bands is plotted in Figure~2a.

\begin{figure}[t!]
\begin{center}
\includegraphics[bb=18 144 592 576,height=9cm]{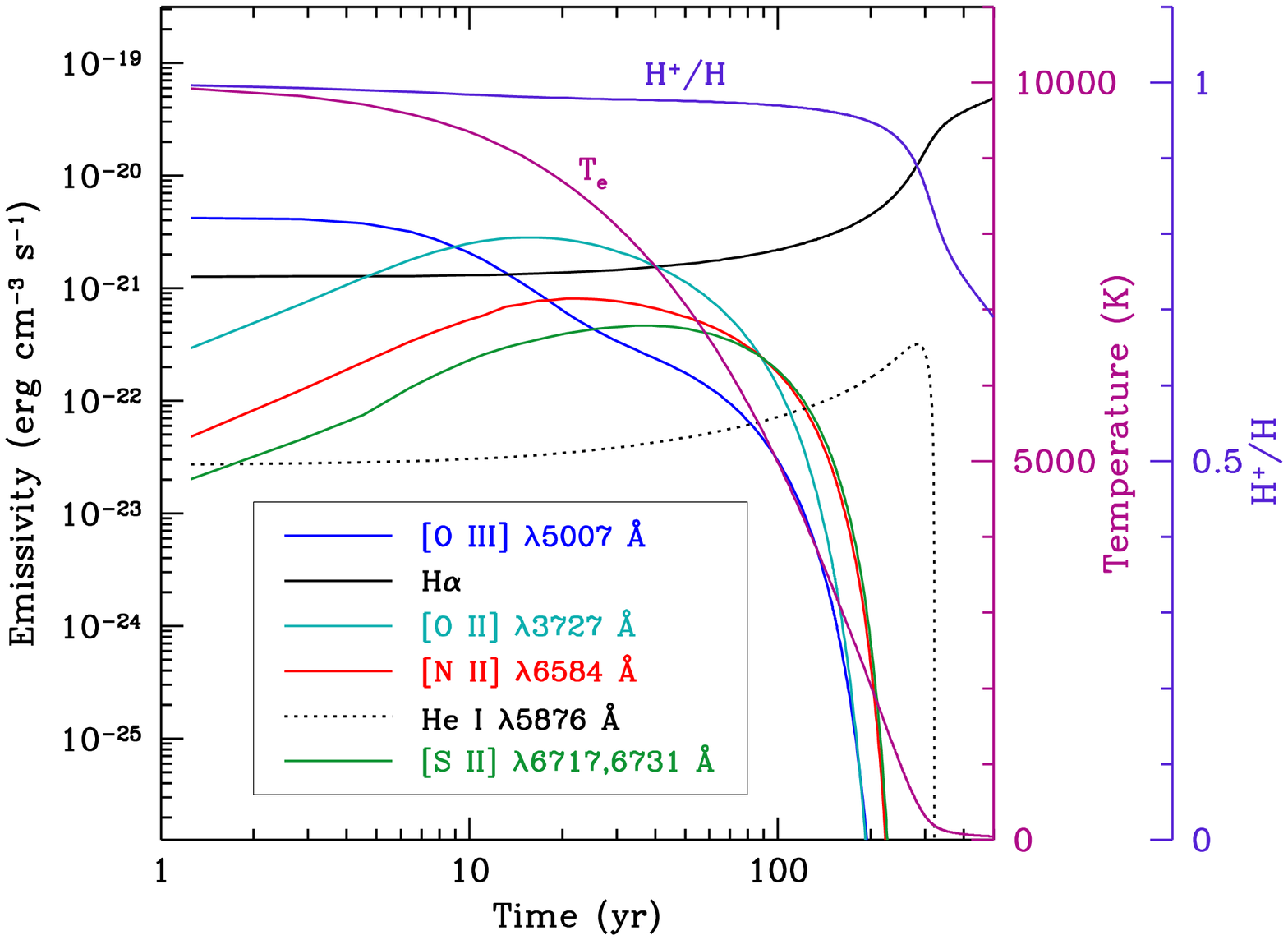}
\end{center}
\vspace*{-0.5cm}
{\small
{\bf
Supplementary Figure 4.--} 
Time-evolution of the electron temperature $T_e$, hydrogen ionization 
fraction H$^+$/H, and emissivity of the H$\beta$, He~{\sc i} $\lambda$5876, 
[O~{\sc ii}] $\lambda$3727, [O~{\sc iii}] $\lambda$5007,
[N~{\sc ii}] $\lambda$6584, and [S~{\sc ii}] $\lambda\lambda$6716,6731 
emission lines of a photo-ionized cloud of gas with solar abundances, 
electron density of 200 cm$^{-3}$ and electron temperature of 10,000 K after 
the ionizing source is switched off at time t=0. 
}
\vspace*{-0.55cm}
\end{figure}

\noindent
{\bf Spectral models of the inner and outer shells of HuBi\,1. } 
The NOT ALFOSC two-dimensional spectra were also used to extract
one-dimensional spectra of the inner and outer shells of HuBi\,1
(Supplementary Figure~3), using the apertures shown in panels 
\emph{b}, \emph{c}, and \emph{d} of Figure~1.  
A region of radius 0\farcs7 at the location of the central 
star was excluded to excise its emission from the spectrum 
of the inner shell.

\noindent
The outer shell is dominated by H~{\sc i} and He~{\sc i} recombination 
lines, with much fainter [N~{\sc ii}] $\lambda\lambda$6548,6584  
and [S~{\sc ii}] $\lambda\lambda$6716,6731, and no [O~{\sc ii}] 
$\lambda$3727 nor [O~{\sc iii}] $\lambda\lambda$4959,5007 
emission lines. 
The density-sensitive [S~{\sc ii}] $\lambda\lambda$6716,6731 doublet 
has been used to derive an electronic density $\approx$200 cm$^{-3}$.
At the distance of HuBi\,1 and adopting a volume filling factor
of 0.3\cite{FPB2016}, this density implies an ionized mass of 0.08
$M_\odot$, which is reasonable value for a PN.    
The code MAPPINGS V 5.1.13\cite{SD2017} was used to explore the evolution
of the intensity of various emission lines in a freely cooling gas. 
The Supplementary Figure~4 shows the time-evolution of the electronic 
temperature, ionization fraction of hydrogen, and emissivity of several 
emission lines for a spherical photo-ionized cloud of gas once the 
ionizing source is switched off and the gas is allowed to cool freely. 
The abundances have been assumed to be solar and the electron temperature 
to be 10,000~K.  
The electron density in the model is 200 cm$^{-3}$, as derived from the 
density-sensitive [S~{\sc ii}] $\lambda\lambda$6717,6731 doublet line ratio.  
As time goes by, O$^{++}$, the most efficient coolant, is the first 
ion to fully recombine, followed by O$^+$, and then by N$^+$ and S$^+$. 
It takes longer for He$^{+}$ to recombine, whereas a significant 
fraction of hydrogen will remain ionized for a much longer period 
of time. 
The outer shell of HuBi\,1 is found to be at the stage when the emissivities 
of the [N~{\sc ii}] and [S~{\sc ii}] emission lines are much higher than that 
of [O~{\sc ii}]. 
In Supplementary Figure~4, this occurs between 90 and 220 years 
after the central source is switched off, but we note that, whereas 
the overall behavior of the gas remains the same, independently 
of the gas density (but for cases of extremely high metallicities), 
the time estimate may vary significantly depending on the metallicity 
of the gas and its density. 
The mild axisymmetry of the outer shell of HuBi\,1 (see the description 
of the morpho-kinematical model) is not expected to
affect significantly the main results presented here.
Furthermore, these effects are certainly mitigated by the
particular orientation of the long-slit, mostly orthogonal
to the outer shell symmetry axis.

\noindent
As for the inner shell, the brightest emission lines in its spectrum 
are those of [N~{\sc ii}]. 
Fainter emission is detected in the forbidden lines of [O~{\sc i}], 
[O~{\sc ii}], [O~{\sc iii}], [N~{\sc i}], and [S~{\sc ii}], as well 
as in the recombination emission lines of H~{\sc i} and He~{\sc i}, 
and the notable detection of the He~{\sc ii} $\lambda$4686 emission 
line.
The presence of high ionization lines in the inner shell, together
with its intricate profile structure, can be attributed to shocks,
as the central source is no longer capable to appreciably ionize
the gas and produce He$^{++}$.  
To explore this possibility, we used also MAPPINGS to evaluate a shock 
expanding through a fully ionized (H$^+$/H=1, He$^+$/He=1) medium with 
solar abundances and a pre-shock density 200 cm$^{-3}$. 
The magnetic field was assumed to have a low value, $B_0=0-1 \mu$G.  
Shocks with velocities above 50 km~s$^{-1}$ are required to produce 
[O~{\sc iii}] emission, and above 70 km~s$^{-1}$ to produce the
observed value of the He~{\sc ii}/H$\beta$ line ratio.  
Much higher shock velocities, above 100 km~s$^{-1}$, can be 
firmly rejected as they imply line ratios very different from 
the ones observed. 
The emission structure behind a 70 km~s$^{-1}$ shock is qualitatively 
consistent with that observed in the inner shell of HuBi\,1, with the 
He~{\sc ii} emission peaking outside at the shock region and [N~{\sc ii}] 
and [O~{\sc iii}] peaking inside in the post-shock cooling region.

\noindent
{\bf Low-mass Very Late Thermal Pulse Models. }
The evolved PN around IRAS\,17514 and the strongly H-deficient composition 
of its stellar wind suggest a very late thermal pulse (VLTP) origin.  
In addition, the slow evolution of the central star implies a low-mass star. 
To test these assumptions, we computed several evolution sequences of VLTP 
events in low-mass stars using LPCODE, La Plata stellar evolution code. 
LPCODE is a one-dimensional stellar evolution code widely used for the 
computation of full evolutionary sequences from the zero age main sequence 
to the WD stage\cite{ASPetal2005}.
%
%
The last version of LPCODE has been carefully calibrated at different 
evolutionary stages to reproduce several AGB and post-AGB observables 
\cite{M2016}.  
LPCODE has been successfully used to compute the evolution of post-AGB 
stars\cite{M2016} and the formation of H-deficient stars through late 
helium flashes\cite{BMCetal2018} 

\begin{figure}[t!]
\begin{center}
\includegraphics[height=15.5cm]{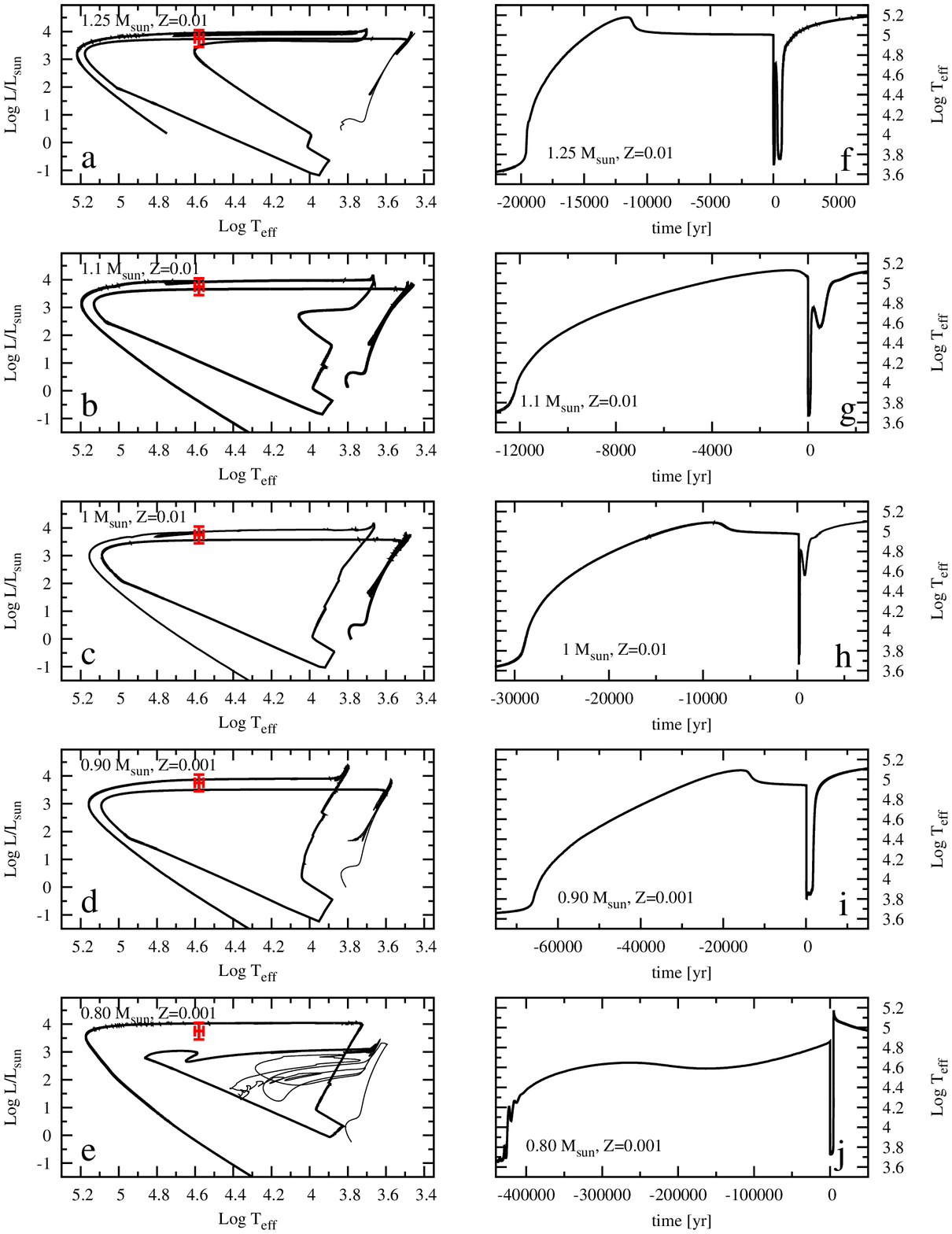}
\end{center}
\vspace*{-0.5cm}
{\small
{\bf
Supplementary Figure 5.-- VLTP sequences of low-mass stars. }  
(left) Evolution of the stellar sequences in the HR diagram, where the
red-cross indicates the location of the central star of HuBi\,1 with 
uncertainties as inferred from our best-fit non-LTE model (see the 
details of the non-LTE model fit).  
(right) Effective temperature evolution of the same sequences during
the departure from the AGB and after the VLTP event (set at t=0). 
}
\vspace*{-0.55cm}
\end{figure}

\noindent
In order to create models of [WC] stars, we have adopted some low-mass
AGB models previously computed\cite{M2016} and adjusted the stellar 
winds during the last stage of the AGB evolution to ensure they left 
the AGB at the right time for a VLTP to take place in the post-AGB 
evolution of the sequence. 
The models correspond to initial masses 0.8 $M_\odot$ and 0.9 $M_\odot$ 
($Z=0.001$) and 1 $M_\odot$, 1.1 $M_\odot$, and 1.25 $M_\odot$ ($Z=0.01$).  
The evolution of these sequences in the HR diagram is shown in the left 
panels of the Supplementary Figure 5, together with our luminosity and 
temperature determinations for IRAS\,17514. 
The location of IRAS\,17514 is consistent with a post-VLTP stage in all 
sequences. 
The surface abundances predicted for these VLTP models are shown in Table~1, 
together with our determinations for the surface abundances of IRAS\,17514. 
The qualitative agreement between the abundances measured
in IRAS\,17514 and the predictions of VLTP computations is 
remarkable, supporting a VLTP event in IRAS\,17514.

\begin{table}[ht!]
\begin{center}
{\bf Table 1.--} Chemical abundances of IRAS\,17514 and those of several VLTP sequences. 
\begin{tabular}{rcccccccc}
\hline
\hline
Initial Mass       &  H         & $^4$He        & $^{12}$C & $^{13}$C & $^{14}$N & $^{16}$O & $^{20}$Ne & $^{22}$Ne \\
\hline
IRAS\,17514        & 0.01--0.05 & 0.33$\pm$0.10 & \multicolumn{2}{c}{0.50$\pm$0.10} & 0.01 & 0.1 & \multicolumn{2}{c}{0.04} \\
\hline
  0.8 $M_\odot$~~~~ & 0.030             & 0.52 & 0.30 & 0.041 & 0.016 & 0.086 & 9$\times$10$^{-5}$ & 1$\times$10$^{-3}$ \\
  0.9 $M_\odot$~~~~ & 5$\times$10$^{-3}$ & 0.48 & 0.29 & 0.069 & 0.065 & 0.091 & 9$\times$10$^{-5}$ & 9$\times$10$^{-4}$ \\
  1.0 $M_\odot$~~~~ & 5$\times$10$^{-5}$ & 0.40 & 0.34 & 0.074 & 0.056 & 0.116 & 9$\times$10$^{-4}$ & 0.01 \\
  1.1 $M_\odot$~~~~ & 5$\times$10$^{-3}$ & 0.38 & 0.37 & 0.059 & 0.028 & 0.145 & 9$\times$10$^{-4}$ & 0.01 \\
 1.25 $M_\odot$~~~~ & 5$\times$10$^{-6}$ & 0.41 & 0.33 & 0.067 & 0.047 & 0.133 & 9$\times$10$^{-4}$ & 0.01 \\
\hline
\hline
\end{tabular}
\end{center}
\end{table}

\noindent
These models should also reproduce the time behavior of IRAS\,17514. 
For this purpose, we show in the right panels of the Supplementary Figure~5 
the time evolution of $T_{eff}$ from the departure from the AGB ($\log T_{eff} 
< 3.7$) to the beginning of the WD phase ($\log T_{eff} \sim 5$) and through 
the VLTP event (chosen as t=0).  
The low density and morphological appearance of HuBi\,1 suggest it
is an evolved PN, which is further supported by its kinematical age
$\approx$9,000 yr. 
On the other hand, the outer nebula is recombining, indicating that the 
ionizing source faded 90-220 years ago, whereas spectroscopic observations 
of IRAS\,17514 show that its temperature has stayed almost constant at 
$T_{\star} \simeq 38,000$ K for at least two decades.

\noindent
The 0.8 $M_\odot$ and 0.9 $M_\odot$ sequences take $\approx$400,000 and
$\approx$60,000 years, 
respectively, between the departure from the AGB and the VLTP event, which is 
inconsistent with the estimated age of HuBi\,1 (and even with the accepted 
visibility lifetime of PNe, $\approx$20,000 yr). 
Consequently, the evolution before the VLTP event can be used 
to reject very low-mass VLTP events.  
On the contrary, the time between the departure from the AGB and the point 
of maximum $T_{eff}$ of the 1.0 $M_\odot$, 1.1 $M_\odot$, and 1.25 $M_\odot$ 
sequences is still below the accepted visibility lifetimes of PNe.

\noindent
After the VLTP event, the 1.25 $M_\odot$ sequence stays as a born-again AGB 
giant for about 100 yr, then it reheats to 38,000 K, makes a double loop, 
crosses $T_{eff}$ at 38,000 K again, stays as a giant for $\approx$150 years,
and about 700 years after the star reheats back to 38,000~K. 
This sequence stays within the $T_{eff} \simeq 36,000-40,000$ K range
for only 1, 7, and 5 years in each crossing, respectively, and thus 
it does not reproduce the observed spectral behavior of IRAS\,17514 
between 1989 and 2017.
As for the 1.0 $M_\odot$ sequence, it is able to describe the behavior of 
IRAS\,17514 in this period of time, as well as the existence of an old 
evolved PN, but the time-lapse between the departure from the AGB and the 
VLTP event for our 1.0 $M_\odot$ sequence cannot be shortened below 
$\simeq$22,000 years, resulting in a disagreement of a factor of two with 
the inferred kinematical age of the PN. 
Therefore, the VLTP evolution can be used to reject the 
1.0 $M_\odot$ and 1.25 $M_\odot$ sequences, with the 1.1 
$M_\odot$ sequence being the only one capable to reproduce 
accurately the time behavior of IRAS\,17514 and its nebula.

\noindent
As shown in Figure 3, the 1.1 $M_\odot$ sequence departs from the AGB about 
$\approx$12,000 yr before the occurrence of the VLTP event, in reasonable 
agreement with the kinematical age of HuBi\,1. 
We note, however, that the post-VLTP evolution depends on the details 
of the mass removed by the stellar winds during the return to the AGB 
as a H-deficient giant (i.e., the Born Again AGB phase), which is very 
uncertain. 
Observations of bona fide VLTP events (V605\,Aql, V4334\,Sgr) 
indicate very high mass-loss rates, in the range 
$10^{-3}-10^{-5} M_\odot$~yr$^{-1}$\cite{HZHetal2005,CBLetal2013}, 
once the star becomes a cold AGB giant\cite{DLSetal2000} ($\log T_{eff} 
< 3.8$).  
Consequently, we performed simulations of the post-VLTP evolution of our 
1.1 $M_\odot$ sequence under different assumptions for the mass removed 
at $\log T_{eff} < 3.8$ during this phase. 
The Supplementary Figure 6 shows the details of such simulations. 
In particular, a mean mass-loss rate of 7.6$\times$10$^{-5} M_\odot$~yr$^{-1}$,
i.e., within the observed range in other born-again events, allows us to
reproduce the observed behavior of IRAS\,17514 between 1989 and 2017. 
The mass of the post-AGB star before the VLTP event is 
0.551 $M_\odot$ and the total mass ejected amounts to 
8$\times$10$^{-4}$ $M_\odot$.

\begin{figure}[t!]
\begin{center}
\includegraphics[height=7cm]{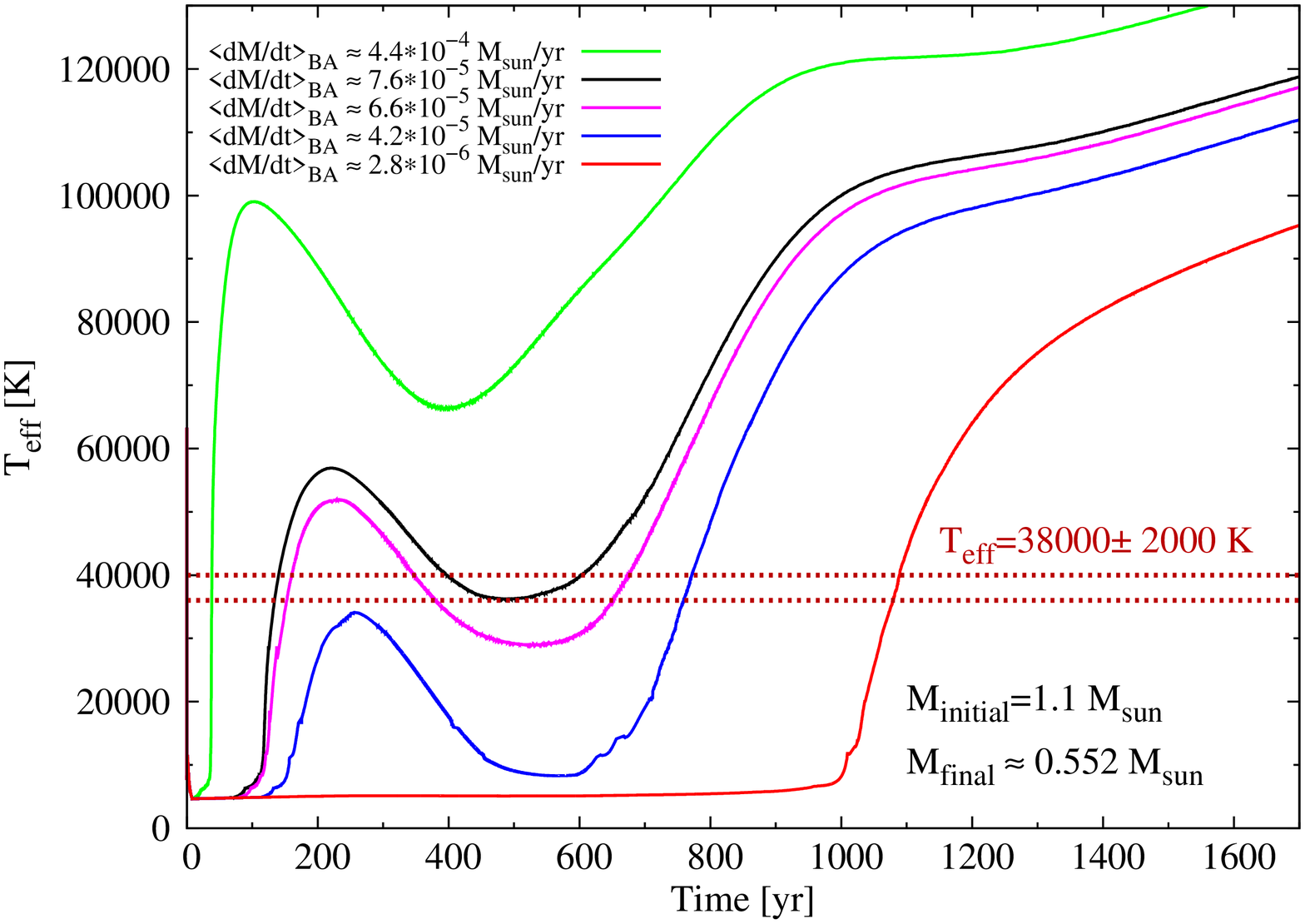}
\end{center}
\vspace*{-0.5cm}
{\small
{\bf
Supplementary Figure 6.-- } 
Predicted $T_{eff}$ evolution of our 1.1 M$_\odot$ sequence after the VLTP 
event under different assumptions of the mass-loss rate through winds during 
the born-again AGB phase. 
The legend indicates the mean envelope mass removed per year at 
$\log T_{eff} < 3.8$. 
}
\vspace*{-0.55cm}
\end{figure}

\noindent
The time evolution of this sequence is shown in Figure 3. 
This sequence matches most observed properties of HuBi\,1 and its central star. 
It explains the existence of an old PN with kinematical age $\approx$9,000 yr, 
and the almost constant temperature $T_{\star} = 38000\pm2000$ K of the CSPN in 
the period of time between 1996 and 2014, its chemical enrichment, and its 
stellar luminosity $\log(L/L_\odot) =3.75\pm0.3$.

\noindent
{\bf Acnowledgements}

\noindent
Some of the data presented in this article were obtained with ALFOSC, 
which is provided by the Instituto de Astrof\'\i sica de Andaluc\'\i a 
(IAA) under a joint agreement with the University of Copenhagen and 
NOTSA.  
The Nordic Optical telescope (NOT) is installed in the Spanish Observatorio 
del Roque de los Muchachos of the Instituto de Astrof\'\i sica de Canarias, 
in the island of La Palma (Spain). 
This article is also based upon observations carried out at the Observatorio 
Astron\'omico Nacional on the Sierra San Pedro M\'artir (OAN SPM), Baja 
California, Mexico. 
We thank the daytime and night support staff at the OAN 
SPM for facilitating and helping obtain our observations. 
A.A.\ and C.M.\ acknowledge support through the CONACyT project 
CONACyT-CB2015-254132.  
G.R.-L.\ acknowledges support from Universidad de Guadalajara, Fundaci\'on 
Marcos Moshinsky, ProMoFID2018 and CONACyT (grant A1-S-12258).
L.S.\ acknowledges support from PAPIIT grant IA-101316 (Mexico). 
L.F.M.\ is supported by Ministerio de Econom\'\i a, Industria y Competitividad 
(Spain) grants AYA2014-57369-C3-3 and AYA2017-84390-C2-1-R (cofunded by FEDER 
funds). 
M.A.G.\ acknowledges support of the grant AYA 2014-57280-P, cofunded with 
FEDER funds.
M.M.M.B.\ is partially supported through ANPCyT grant PICT-2016-0053
and MinCyT-DAAD bilateral cooperation program through grant DA/16/07.  
S.A.Z.\ was supported by the ITE-UNAM agreement 1500-479-3-V-04.

\noindent
{\bf Author Contribution. }

\noindent
M.A.G.\ planned the research project, programmed the observations, wrote the 
main body of the manuscript, and organized the writing of some subsections.  
A.A.\ performed the MAPPINGS simulations, C.M.\ the CLOUDY ones, and 
both devised the excitation nature of the inner and outer shells.
C.K.\ estimated the ionizing flux necessary for the inner and outer shells.  
G.R.-L.\ reduced the imaging data and contributed to the analysis of the
time-evolution of the central star.
H.T.\ analised the spectrum of the central star using PoWR to
determine its stellar parameters and abundances.  
L.S.\ and G.R.-L.\ obtained and reduced the high-dispersion spectroscopic 
observations, and together with L.F.M.\ and S.A.Z.\ analysed them using 
SHAPE.  
M.M.M.B.\ devised the post-AGB evolutionary scenario and computed the
LPCODE VLTP evolutionary sequences.
X.F.\ reduced the spectroscopic data and carried out the analysis of
one-dimensional spectra and spatial profiles of emission lines.  
All authors contributed to the discussion of the different sections
of this work.

\noindent
{\bf Data and code availability.} 

\noindent
The data that support the plots and other findings in this paper are 
available from the corresponding author on a reasonable request.
CLOUDY, MAPPINGS and SHAPE can be freely downloaded from
https://www.nublado.org, 
https://mappings.anu.edu.au/code and
http://www.astrosen.unam.mx/shape/index.html, 
respectively.
The LPCODE and PoWR codes used in this paper are similarly available 
under request from M.M.\ Miller Bertolami and H.\ Todt, respectively.

\end{document}